\documentclass[
 reprint,
 amsmath,amssymb,superscriptaddress,
 aps,
 prl,
]{revtex4-1}

\usepackage{graphicx}
\usepackage{dcolumn}
\usepackage{bm}
\usepackage{color}

\begin{document}

\title{Fermi Contour Anisotropy of GaAs Electron-Flux Composite Fermions in Parallel Magnetic Fields}

\author{D.\ Kamburov}
\author{M.~A.\ Mueed}
\author{M.\ Shayegan}
\author{L.~N.\ Pfeiffer}
\author{K.~W.\ West}
\author{K.~W.\ Baldwin}
\author{J.~J.~D.\ Lee}
\affiliation{ Department of Electrical Engineering, Princeton University, Princeton, New Jersey 08544, USA}
\author{R.\ Winkler}
\affiliation{Department of Physics, Northern Illinois University, DeKalb, Illinois 60115, USA}
\affiliation{Materials Science Division, Argonne National Laboratory, Argonne, Illinois 60439, USA}

\date{\today}

\begin{abstract}
In high-quality two-dimensional electrons confined to GaAs quantum wells, near Landau level filling factors $\nu=$ 1/2 and 1/4, we observe signatures of the commensurability of the electron-flux composite fermion cyclotron orbits with a unidirectional periodic density modulation. Focusing on the data near $\nu=1/2$, we directly and quantitatively probe the shape of the composite fermions' cyclotron orbit, and therefore their Fermi contour, as a function of magnetic field ($B_{||}$) applied parallel to the sample plane. The composite fermion Fermi contour becomes severely distorted with increasing $B_{||}$ and appears to be elliptical, in sharp contrast to the electron Fermi contour which splits as the system becomes bilayer-like at high $B_{||}$. We present a simple, qualitative model to interpret our findings.
\end{abstract}


\maketitle

In the presence of a strong perpendicular magnetic field ($B_{\perp}$) and at very low temperatures, high-quality two-dimensional (2D) electron systems exhibit the fractional quantum Hall effect (FQHE), described elegantly by the concept of composite fermions (CFs) \cite{Jain.2007, Jain.PRL.1989, Halperin.PRB.1993}. In this formalism, CFs are quasi-particles formed by the attachment of an even number of flux quanta to each electron in high $B_{\perp}$. At even-denominator Landau level filling factors, e.g. at $\nu=$ 1/2 or 1/4, the flux attachment completely cancels the external field, leaving the CFs as if they are at zero \textit{effective} magnetic field. Analogous to electrons at low fields, the CFs near these fillings occupy a Fermi sea with a well-defined Fermi contour \cite{Jain.2007, Jain.PRL.1989, Halperin.PRB.1993, Willett.PRL.1993, Kang.PRL.1993, Goldman.PRL.1994, Smet.PRL.1996}.

The problem of anisotropy in FQHE phenomena has sparked recent interest both experimentally and theoretically \cite{Balagurov.PRB.2000, Gokmen.Nature.2010, Xia.Nat.Phys.2011,  Mulligan.PRB.2010, Yang.Haldane.PRB.2012, Qui.Haldane.PRB.2012,  Wang.PRB.2012, Yang.Cond.Mat.2013, Z.Papic.2013}. In particular, the existence of a CF Fermi contour raises the question whether this contour is anisotropic if the low-field particles have an anisotropic Fermi contour \cite{Balagurov.PRB.2000, Gokmen.Nature.2010}. This issue was addressed very recently for \textit{hole-flux} CFs \cite{Kamburov.CFs.PRL.2012}. It was established that a parallel magnetic field $B_{||}$ which induces anisotropy in the hole Fermi contour also makes the CF Fermi contour anisotropic, although the observed anisotropy is much smaller for the CFs. The reason for the Fermi contour anisotropy in both the hole and hole-flux CFs cases is the finite (non-zero) thickness of the quasi-2D system and the coupling of the carriers' out-of-plane motion to $B_{||}$. While the Fermi contour anisotropy for the quasi-2D holes is semi-quantitatively understood from energy band calculations \cite{Kamburov.PRB.86.2012}, there are no theoretical predictions for the anisotropy expected for hole-flux CFs. Calculations for such anisotropy would be particularly challenging because of the complex structure of 2D hole Landau levels, especially in tilted magnetic fields \cite{Winkler.Book.2003}.

Here we report direct and quantitative measurements of CF Fermi contour anisotropy induced by $B_{||}$ in 2D \textit{electron} systems confined to GaAs quantum wells (QWs) with widths ranging from 30 to 50 nm. For a given QW width, the Fermi contour anisotropy and shape for the CFs is drastically different from their electron counterparts. We also demonstrate that the degree of anisotropy critically depends on the QW width which determines the thickness of the electron layer. Our quantitative data, together with the much simpler band structure of GaAs electrons should stimulate efforts to theoretically explain the $B_{||}$-induced CF Fermi contour anisotropy.

We studied 2D electrons confined to symmetric GaAs QWs grown via molecular beam epitaxy on (001) GaAs substrates. We measured three samples with QW widths of 30, 40, and 50 nm, located 190 nm under the surface, and flanked on each side by 95-nm-thick, undoped Al$_{0.24}$Ga$_{0.76}$As barrier layers and Si $\delta$-doped layers. The 2D electron densities $n$ in our samples are $\simeq 1.8 \times 10^{11}$ cm$^{-2}$, and the mobilities are $\simeq10^7$ cm$^2$/Vs. We did experiments in two $^3$He refrigerators with base temperatures of $T\simeq$ 0.3 K and with an 18 T superconducting or a 31 T resistive magnet.

\begin{figure}[t!]
\includegraphics[trim=0cm 0cm 0cm 1.2cm, clip=false, width=.48\textwidth]{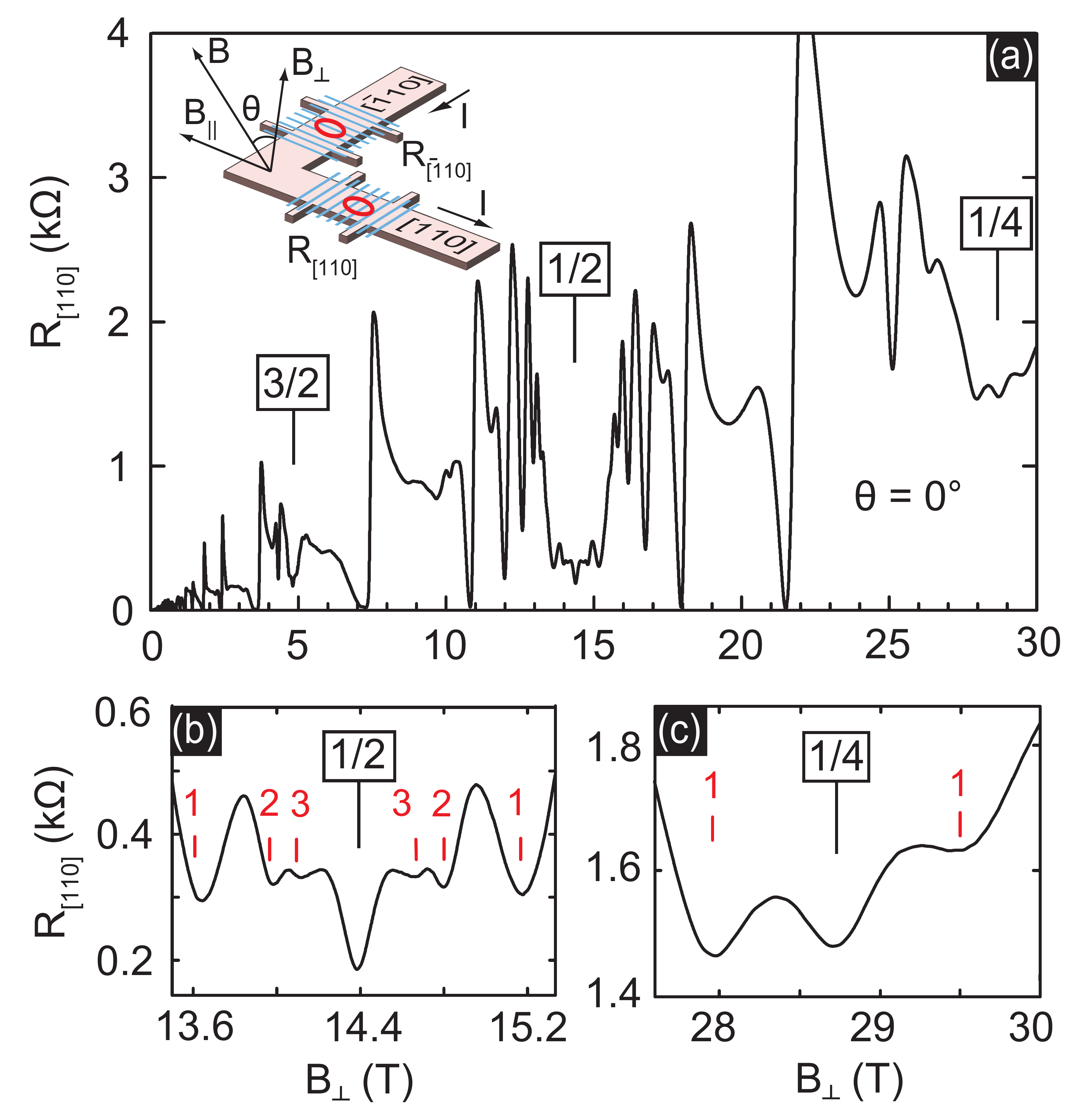}
\caption{\label{fig:Fig1} (color online) Inset: an L-shaped Hall
bar with periodic superlattices of negative electron-beam resist. (a) Magneto-resistance trace from the [$110$] Hall bar of the 40-nm-wide QW sample with $n=1.78 \times 10^{11}$ cm$^{-2}$. (b), (c) Prominent commensurability resistance minima are seen near $\nu=1/2$ and 1/4. The positions of the resistance minima expected for fully spin-polarized CFs with a circular Fermi contour are marked with indexed vertical lines (see text). }
\end{figure}

Our technique for measuring the CF Fermi contour anisotropy is illustrated in Fig.\;\ref{fig:Fig1}(a) inset. Each sample has two Hall bars, oriented along the [$110$] and [$\overline{1}10$] directions. The Hall bars are covered with periodic gratings of negative electron-beam resist with a period $a=200$ nm. The stripes impart a periodic strain to the sample surface which, through the piezoelectric effect in GaAs, induces a periodic density modulation \cite{Skuras.APL.1997, Endo.PRB.2000, Endo.PRB.2001, Endo.PRB.2005, Kamburov.PRB.2012b,  Kamburov.2012b, Kamburov.CFs.2012}. In the presence of an applied $B_{\perp}$, whenever the diameter of the charged carriers' cyclotron orbit becomes commensurate with the period of the density modulation, the sample's resistance exhibits a minimum. The anisotropy of the cyclotron orbit can therefore be simply determined via measuring the positions of the commensurability magneto-resistance minima along the two perpendicular arms of the L-shaped Hall. Since the reciprocal-space ($k$-space) orbits are expected to be a scaled version of the real-space trajectories, rotated by 90$^{\circ}$ \cite{AM}, our commensurability measurements then directly probe the Fermi contour shape. In our experiments, we tilted the sample around the [$\overline{1}10$] direction so that $B_{||}$ was always along [$110$], with $\theta$ denoting the angle between the field direction and the normal to the 2D plane (Fig. 1(a) inset). Note that, in quasi-2D systems where the carriers have a finite layer thickness, the application of $B_{||}$ shrinks the real-space cyclotron orbit diameter in the in-plane direction perpendicular to $B_{||}$; the Fermi contour then should shrink in the direction of $B_{||}$.

\begin{figure}[t]
\includegraphics[trim=0.1cm 0.4cm 0cm 0.3cm, clip=true, width=0.48\textwidth]{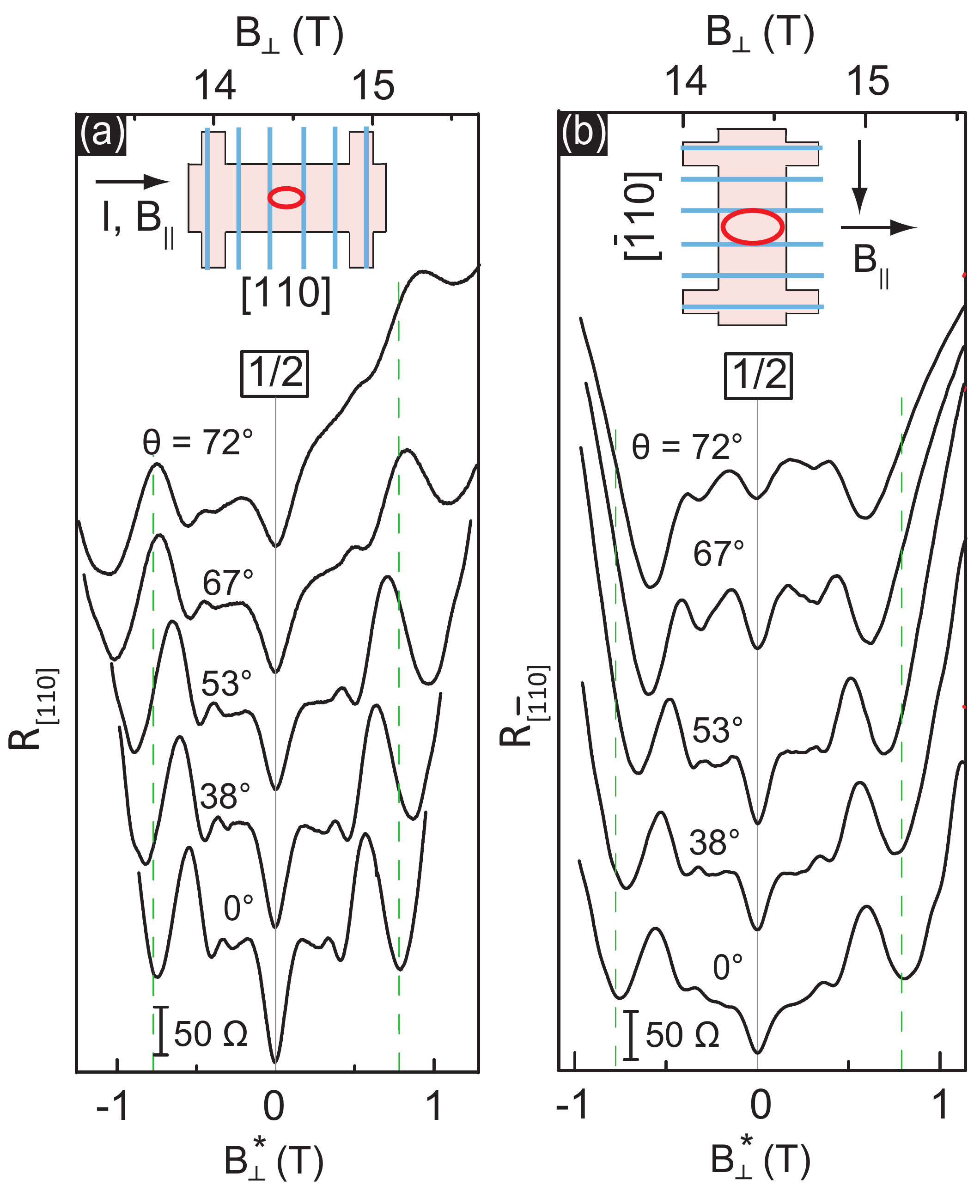}
\caption{\label{fig:Fig2} (color online) Evolution of the
magneto-resistance near $\nu=1/2$ for the 40-nm-wide QW sample measured along the [$110$] and [$\overline{1}10$] Hall bars. The tilt angle $\theta$ is given for each trace, and the traces are shifted vertically for clarity. The vertical green dashed lines mark the expected positions of the primary commensurability resistance minima if the CF cyclotron orbit were circular. In both panels, the scale for the applied \textit{external} field $B_{\perp}$ is shown on top while the bottom scale is the \textit{effective} magnetic field $B^*_{\perp}$ felt by the CFs.}
\end{figure}

In Fig.\;\ref{fig:Fig1} we show the magneto-resistance trace at $\theta=0^{\circ}$ ($B_{||}=0$) along the [110] Hall bar of the 40-nm-wide QW sample. It exhibits prominent commensurability features near $\nu=1/2$ and $1/4$, including a characteristic, V-shaped, resistance dip at $\nu=1/2$, followed by several resistance minima on each side of $\nu=1/2$ (marked by $i=1, 2, 3$ in Fig.\;\ref{fig:Fig1}(b)) and flanked by regions of rapidly rising resistance. The positions of the resistance minima follow closely those expected from the \textit{magnetic} commensurability condition for \textit{spin-polarized} CFs with \textit{circular} Fermi contour, namely $2R_C^*/a=i+1/4$, where $i=1,2,3,...$; $2R_C^*=2 \hbar k_F^*/eB^*_{\perp}$ is the CF cyclotron orbit diameter, $k_F^*=\sqrt{4\pi n}$ is the CF Fermi wave vector, and $B^*_{\perp}=B_{\perp}-B_{\perp,1/2}$ is the effective field seen by the CFs near $\nu=1/2$ \cite{footnote101}. This is consistent with previous reports of commensurability features for electron-flux CFs near $\nu=1/2$ \cite{Willett.PRL.1997, Smet.PRB.1997, Smet.PRL.1998, Mirlin.PRL.1998, Oppen.PRL.1998, Smet.PRL.1999, Zwerschke.PRL.1999}, except that here we see the additional $i=2$ and 3 minima, attesting to the very high quality of the sample and the periodic potential. In the present study, we monitor the shift in the positions of the $\nu=1/2$ primary CF commensurability minima $(i=1)$ as a function of applied $B_{||}$ to directly probe the size and shape of the CF Fermi contour. Note that commensurability resistance minima are also seen near $\nu=1/4$ at very high fields (Fig.\;\ref{fig:Fig1}(c)); this is the first direct observation of ballistic transport and geometric resonance for \textit{four-flux} CFs. The positions of the resistance minima match the expected values based on a \textit{magnetic, spin-polarized} commensurability condition equivalent to the one near $\nu=1/2$. We also observe commensurability features near $\nu=3/2$, and their positions are consistent with spin-unpolarized $\nu =3/2$ CFs \cite{Endo32.PRB}.

As illustrated in Fig.\;\ref{fig:Fig2}, the application of $B_{||}$ profoundly affects the positions of the commensurability resistance minima near $\nu=1/2$. Data for the two Hall bars along the [$110$] and [$\overline{1}10$] directions are shown in Figs.\;\ref{fig:Fig2}(a) and (b). In both panels, the vertical green dashed lines mark the expected positions of the CF $i=1$ resistance minima based on magnetic condition for spin-polarized CFs with circular Fermi contours. These lines match very well the observed positions of the resistance minima for the bottom traces of Fig.\;\ref{fig:Fig2}, which were taken at $\theta=0$ ($B_{||}=0$). With increasing $\theta$ and $B_{||}$, for the [$110$] Hall bar (Fig.\;\ref{fig:Fig2}(a)), the positions of the two resistance minima shift away from the dashed lines to \textit{higher} values of $|B^*_{\perp}|$. In contrast, the resistance minima for the Hall bar in the perpendicular, [$\overline{1}10$] direction (Fig.\;\ref{fig:Fig2}(b)) move towards \textit{lower} $|B_{\perp}^*|$, and the shift is smaller.

\begin{figure}[t]
\includegraphics[trim=0.5cm 0.1cm 0.0cm 0.0cm, clip=true, width=0.48\textwidth]{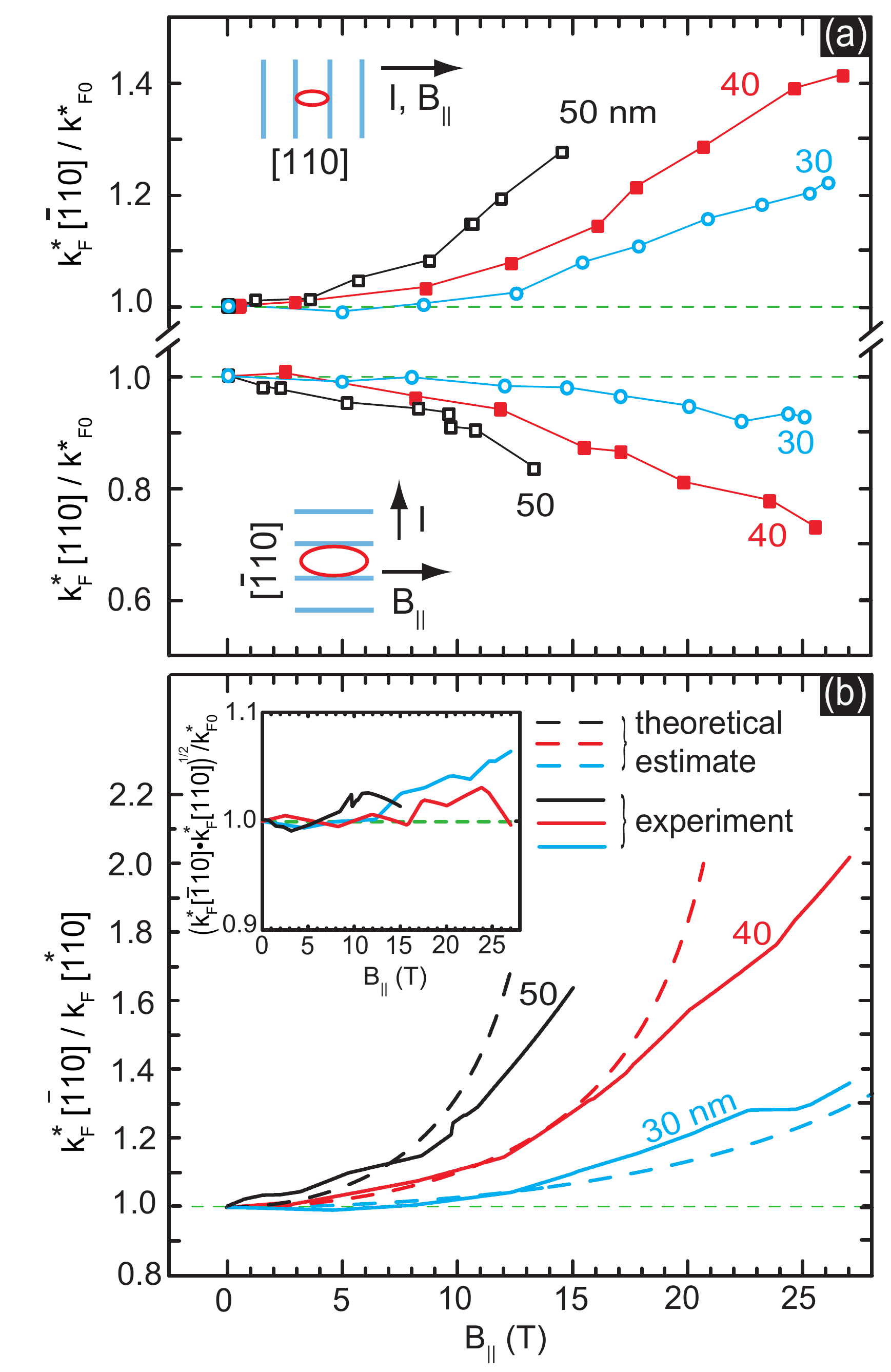}
\caption{\label{fig:Fig3} (color online) (a) Measured values of
the CF Fermi wave vectors $k_F^*$ along the [$\overline{1}10$] and [$110$] directions, normalized to $k_{F0}^*$ (see text). Data shown with open circles (blue), filled squares (red), and open squares (black) are from the 30-, 40-, and 50-nm-wide QWs, respectively. (b) Solid lines: Relative anisotropy of the CF Fermi contours in each QW deduced from dividing the (interpolated) measured values of $k_F^*$ along [$\overline{1}10$] by those along [$110$]. Dashed lines: theoretical estimate of the anisotropy using Eq. (1) (see text). Inset: geometric mean of the measured values of $k^*_F$ along the two directions normalized to $k^*_{F0}$ for each QW.}
\end{figure}

We use the positions of the resistance minima along the [$\overline{1}10$] and [110] directions to directly extract the magnitude of the CF Fermi wave vectors along [110] and [$\overline{1}10$], respectively \cite{footnote104}. From the commensurability condition near $\nu=1/2$ with $i=1$, $k^*_F=(5/8)(eaB^*_{\perp}/\hbar)$. Here $B^*_{\perp}$ indicates the effective CF magnetic field at which the resistance minimum is observed. Using this relation, we converted the $B_{\perp}^*$ positions of the resistivity minima seen in Fig.\;\ref{fig:Fig2} to the size of the CF $k^*_F$ along the [110] and [$\overline{1}10$] directions. The results, normalized to $k_{F0}^*$, the value of $k_F^*$ at $B_{||}=0$, are summarized in Fig.\;\ref{fig:Fig3}(a) for $B^*_{\perp}>0$; the data $B^*_{\perp}<0$ are similar. Clearly, with increasing $B_{||}$, the CF Fermi wave vector along [$\overline{1}10$] increases while along [$110$] it decreases. The data indicate a severe $B_{||}$-induced anisotropy of the CF Fermi contour, up to a factor of $\simeq$ 2 at $B\simeq 25$ T for the 40-nm-wide QW.

\begin{figure}[t]
\includegraphics[trim=0.1cm 0.0cm 0cm 0cm, clip=true, width=0.485\textwidth]{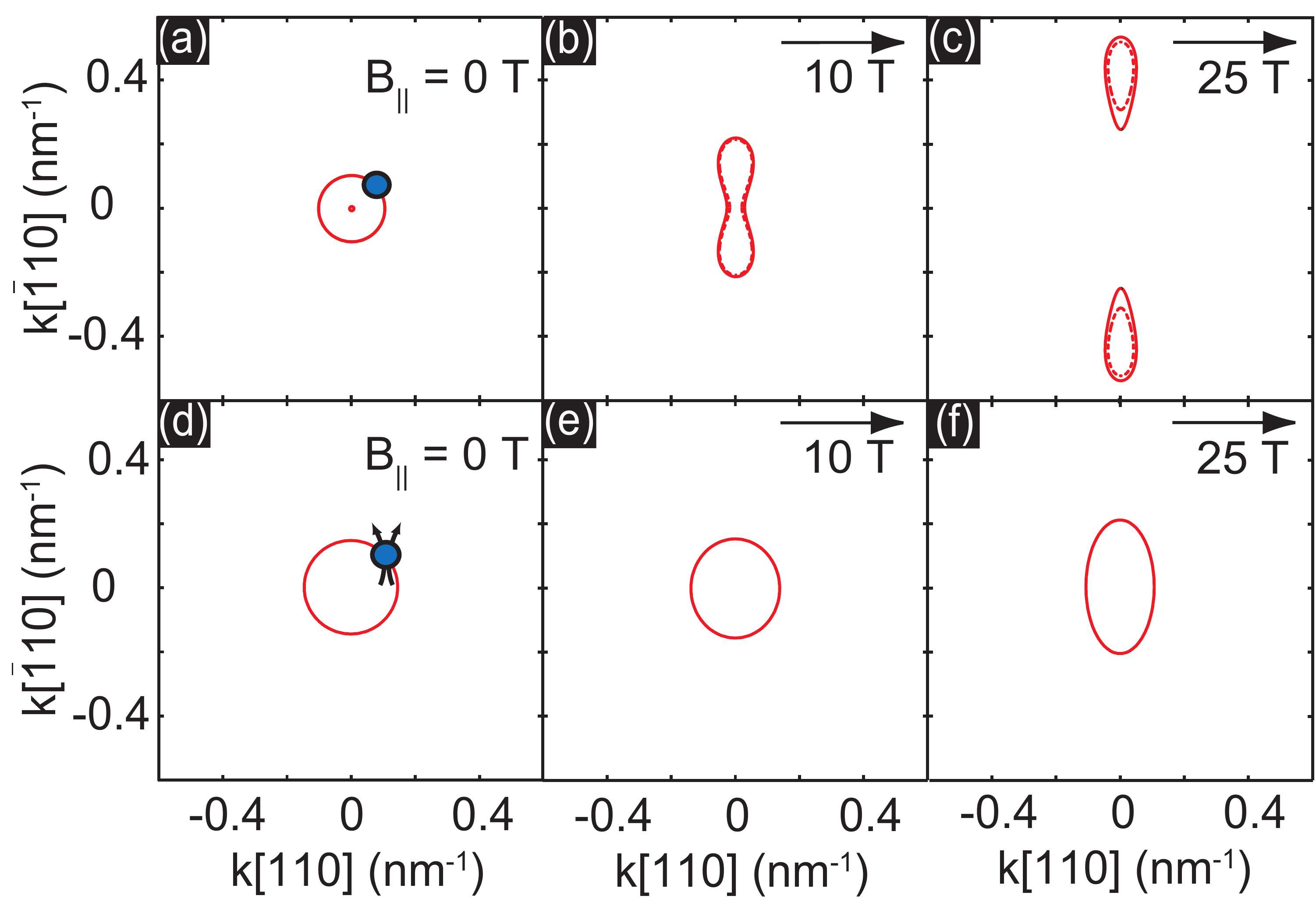}
\caption{\label{fig:Fig4} (color online) (a)-(c) Calculated
electron Fermi contours for a 40-nm-wide GaAs QW with density $n=1.7 \times 10^{11}$ cm$^{-2}$ at $B_{||}=0, 10, 25$ T. The solid and dotted contours correspond to the majority- and minority-spin electrons, respectively. (d)-(f) Evolution of the distortion of the CF Fermi contour with $B_{||}$ measured in the 40-nm-wide QW.}
\end{figure}

To probe the role of the layer thickness on the anisotropy of the CF Fermi contour, we performed similar measurements on the 30- and 50-nm-wide QWs. The results, also summarized in Fig. 3(a), reveal that the wider the QW, the larger the anisotropy. This is best seen in Fig.\;\ref{fig:Fig3}(b) which shows the ratio of the measured $k^*_F$ along [$\overline{1}10$] and [$110$] for each QW. This ratio is $\simeq$ 1.6 at $B_{||}=15$ T for the 50-nm-wide QW, indicating a severe distortion as a result of $B_{||}$. In contrast, the ratio at $B_{||}=15$ T is $\simeq$ 1.3 for the 40-nm-wide QW, and only $\simeq$ 1.1 for the 30-nm-QW.

Next we discuss the shape of the CF Fermi contour. Since in our experiments we measure $k^*_F$ only along two specific, perpendicular directions, we cannot rule out a complicated shape. However, our data are consistent with nearly elliptical CF Fermi contours. As seen in Fig.\;3(b) inset, the geometric mean of the two $k^*_F$'s we measure along [$\overline{1}10$] and [$110$], divided by $k^*_{F0}$, the wave vector expected for a circular CF Fermi contour whose enclosed area is equal to the density of CFs, is close to unity (to better than 7 $\%$) in our entire range of $B_{||}$. This implies that the areas enclosed by elliptical contours with major and minor $k^*_F$'s equal to those we measure indeed enclose the area needed to account for the CFs.

The data presented here provide direct and quantitative evidence that the CF Fermi contours become anisotropic with the application of $B_{||}$. The strong dependence of the distortion on $B_{||}$ and on the QW width implies that the origin of this anisotropy is the coupling between $B_{||}$ and the out-of-plane motion of the CFs. Such coupling is known to severely distort the Fermi contour of \textit{low-field} carriers \cite{Kamburov.preprint.cond.mat, Kamburov.2012b}. Indeed we have experimentally measured the Fermi contours for low-field carriers as a function of $B_{||}$ and the data, which are overall in good agreement with the calculations, show severe Fermi contour distortions. Figure 4 provides comparisons between the Fermi contours for the electrons (Figs. 4(a-c)) in the 40-nm-wide QW sample at $B_{||}=0$, 10, and 25 T, calculated self-consistently based on the 8 $\times$ 8 Kane Hamiltonian \cite{Winkler.Book.2003, Kamburov.preprint.cond.mat}, and the CF Fermi contours at the corresponding $B_{||}$, deduced from our measurements (Figs.\;4 (d-f)). For both electrons and CFs, the Fermi contours become significantly distorted with increasing $B_{||}$, but the distortion is much more severe for the electrons. At sufficiently large $B_{||} (>12$ T), the electrons in fact split into a bilayer-like system with a disconnected contour. Remarkably, however, the CF Fermi contour remains connected up to the highest $B_{||}=25$ T \cite{footnote107}.

We emphasize that, to date, there are no theoretical calculations which treat the anisotropy of CF Fermi contours in the presence of $B_{||}$ \cite{footnote105}. Balagurov and Lozovik \cite{Balagurov.PRB.2000} do not consider the role of $B_{||}$ but rather assume a general 2D system with an anisotropic zero-field Fermi contour. They predict that the CF Fermi contour shape should be identical to that of the zero-field particles; this is in apparent contradiction to our data. Furthermore, besides its thickness, other parameters of the quasi-2D carrier system, such as the details of the band structure and effective mass as well as the character of the Landau level (LL) where the CFs are formed, also play important roles in determining the anisotropy of the CF Fermi contour in a strong $B_{||}$. For example, hole-flux CFs in a 17.5-nm-wide GaAs QW \cite{Kamburov.CFs.2012} exhibit a Fermi contour anisotropy of $\simeq$ 1.2 at $B_{||}=15$ T. This is comparable to the anisotropy we observe for electron-flux CFs in the much wider 30- and 40-nm-wide QW samples (Fig.\;3(b)), indicating that layer thickness is not the only parameter that determines the CF Fermi contour anisotropy. We add that the LLs for GaAs 2D holes are particularly complex because of their non-linear dependence on $B_{\perp}$ \cite{Winkler.Book.2003} and unexpected crossings in tilted magnetic fields \cite{Graninger.PRL.2011,  Kamburov.hole.LL.crossing.2013}. In contrast, 2D electrons in GaAs QWs have a much simpler set of energy band parameters and LL structure. Our experimental data should therefore provide incentive for future theoretical calculations aimed at understanding the $B_{||}$-induced CF Fermi contour anisotropy.

In closing we present a simple, qualitative model to provide an estimate for the expected CF Fermi contour anisotropy. Assuming an effective mass $m_z$ for the out-of-plane, quantum-confined motion and a parabolic dispersion with an effective mass $m_\|$ for the in-plane motion, we can approximately calculate the effect of an in-plane magnetic field $B_x$ on the in-plane motion. For fully spin-polarized electrons confined to a rectangular QW of width $w$ we find in lowest-order perturbation theory \cite{ste68} an elliptic Fermi contour with minor and major radii
\begin{equation}
\label{eq:ellipse}
k_{x,y} = \sqrt{\frac{n}{\pi}} \left( 1 - \frac{2^{10}}{3^5 \pi^6} \frac{e^2 \, B_x^2}{\hbar^2} \frac{w^4 \, m_z}{m_\|} \right)^{\pm 1/4}.
\end{equation}
Here we may expect that the many-particle physics of CFs characterizes the in-plane dynamics of the quasi particles in our experiments. Thus $m_{||}$ should be approximately the effective mass of CFs that contains electron-electron interactions and has been studied previously \cite{Jain.2007,Manoharan.PRL.1994,cfmass}, namely $m_{||}^{\text{CF}} \simeq 1$ (in units of the free-electron mass, $m_e$). The perpendicular motion of the quasi-particles, on the other hand, should reflect the band dynamics which is characterized by the band mass of electrons in GaAs, $m_z = 0.067$ \cite{Winkler.Book.2003}. Different cartesian components of the motion of these quasi-particles thus reflect very different physics. These components get coupled by $B_{||}$. The predictions of this model for the deformation of the Fermi contour in our samples are shown in Fig.\;3(b) with dashed lines. The magnitude of the predicted anisotropy is comparable to the anisotropy seen in the experiment at low values of $B_{||}$, where we expect the model to be valid \cite{holes.2012}. A quantitative explanation of the data, however, requires theoretical calculations of the CF Fermi contours in the presence of $B_{||}$.

\begin{acknowledgments}
We acknowledge support by the NSF (DMR-0904117, DMR-1305691, and ECCS-1001719) for measurements, and the Moore and Keck Foundations and the NSF MRSEC (DMR-0819860) for sample fabrication and characterization. A portion of this work was performed at the National High Magnetic Field Laboratory which is supported by the NSF Cooperative Agreement No. DMR-1157490, the State of Florida and the DOE. Work at Argonne was supported by DOE BES under Contract No. DE-AC02-06CH11357. We thank J.K. Jain, Z. Papic, and Yang Liu for illuminating discussions, and S. Hannahs, E. Palm, T. Murphy, and A. Suslov at NHMFL for valuable help during the measurements. We also thank Tokoyama Corporation for supplying the negative electron-beam resist TEBN-1 used to make the samples.
\end{acknowledgments}


\begin{thebibliography}{99}
\bibitem{footnote101} Throughout this paper we use $^*$ to denote CF quantities.
\end{thebibliography}
\end{document}